\documentclass[12pt]{article}
\usepackage{epsfig}
\usepackage{cite}
\begin{document}

\begin{titlepage}

\hfill FTUV-12-0501

\hfill IFIC/12-30

\vspace{1.5cm}

\begin{center}
\ 
\\
{\bf\large  Confinement, the gluon propagator and the interquark potential  for heavy mesons}
\\
\date{ }
\vskip 0.70cm
 V. Vento
\vskip 0.30cm

{ \it Departamento de F\'{\i}sica Te\'orica -IFIC\\
Universidad de Valencia-CSIC \\
E-46100 Burjassot (Valencia), Spain.} \\ ({\small E-mail:
vicente.vento@uv.es}) 
\end{center}

\vskip 1cm \centerline{\bf Abstract}
The interquark static potential for heavy mesons described by a massive One Gluon Exchange interaction obtained from the propagator of the truncated   Dyson-Schwinger equations does not reproduced the expected Cornell potential. I show that no formulation based on a finite propagator will lead to confinement of quenched QCD. I propose  a mechanism based on a singular nonperturbative  coupling constant  which has the virtue of giving rise to a finite gluon propagator and (almost) linear confinement. The mechanism can be slightly modified to produce the screened potentials of unquenched QCD.

 \vspace{3cm}

\noindent Pacs: 12.38.Gc, 12.38.Lg, 12.39.Pn, 14.40.Pq


\end{titlepage}

\section{Introduction}

\begin{figure}[htb]
\vskip 0.2cm
\includegraphics[scale=0.80]{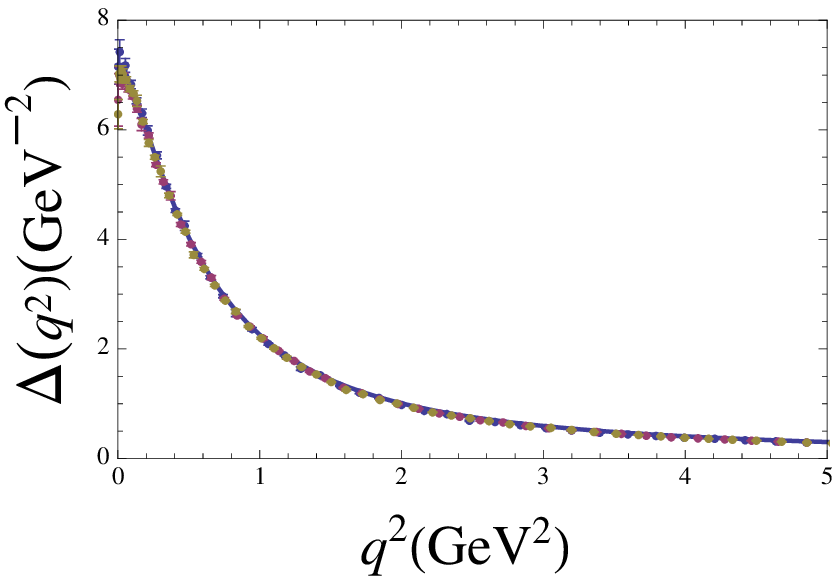}
\vskip -4.6cm \hskip 7.8cm
\includegraphics[scale=0.80]{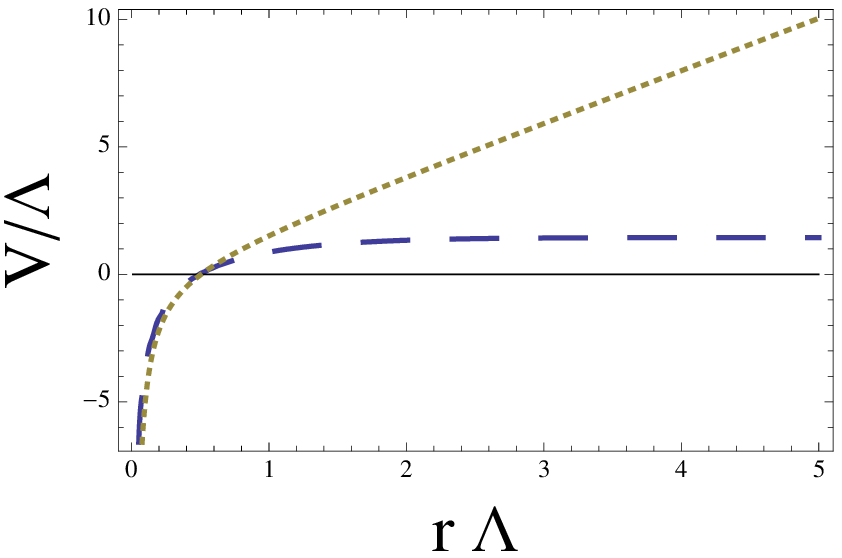}
\vskip 0.2cm
\caption{Left: Fit to the lattice propagator data of ref. \cite{Bogolubsky:2007ud} using the logarithmic mass equation and the following set of parameters for the  Dyson-Schwinger propagator (solid)  $\Lambda = 0.300$ GeV, $\delta = 1/11$, $\rho =2.0$,  $m_0 = 0.374$ GeV, $\mu = 4.5 $ GeV,  $\rho_1= 20.0$, $c= 0.27$. Right: The ``massive" One Gluon exchange potential for the same parameters (long-dashed) \cite{Gonzalez:2011zc} compared with the Cornell potential (dotted).}
\label{OGE}
\end{figure}

The resolution of Dyson-Schwinger equations leads to the freezing of the QCD running coupling (effective charge) in the infrared, which is best understood as a dynamical generation of a gluon mass function, giving rise to a momentum dependence which is free from infrared divergences  \cite{Cornwall:1982zr,Aguilar:2006gr}. Recently, we have calculated the interquark static potential for heavy mesons by assuming that it is given by a massive One Gluon Exchange (OGE) interaction which we have called DS potential \cite{Gonzalez:2011zc}. To our surprise the DS potential does not contain the physics of confinement in the quenched approximation to Quantum Chromodynamics, namely the linear rise at large distances. In  Fig. \ref{OGE}  (right)   the DS potential is shown for the parameters that fit the lattice propagator of ref. \cite{Bogolubsky:2007ud}, using the mass and coupling constant equations of ref. \cite{Gonzalez:2011zc} as  I will recall in section 2, as seen in   Fig. \ref{OGE}  (left).  Note the finiteness of the gluon propagator.

It is well known that a propagator with a functional form $1/q^4$ leads to a linearly rising potential \cite{Gribov:1999ui}, however this propagator is singular at the origin contrary to the  result of the lattice calculation of ref. \cite{Bogolubsky:2007ud}. Finite modifications of the confining Gribov propagator, $\sim 1/(q^2 + m^2)^2$,  have been proposed. These modifications lead to the description of chiral symmetry breaking through confinement with a mass paramenter $\sim 200$ MeV \cite{Cornwall:2010ap}.  Does this modified Gribov (mG) propagator describe confinement in the quenched approximation? It is clear that its funtional form in $r$ space is a decreasing exponential and therefore in some range behaves linearly. Is the parameter space adequate to support linearity?

I proceed by adding the mG propagator  to the DS propagator \cite{Aguilar:2011yb} and study thereafter the corresponding potential. I am able to fit the propagator data quite precisely with this Ansatz. However, the corresponding potential  produces almost no confinement. The value obtained  for the mass paramenter of the mG propagator is too large and the behavior is exponentail even for small values of $r$. If , on the contrary, I fit  the Cornell potential, I get  a Gribov type propagator, i.e. the mass parameter of the mG propagator tends to zero, leading to a huge rise close to the origin in disagreement with lattice QCD.

The corollary of my mathematical analysis of the propagator  data and the Cornell potential  is that a good potential requires a singularity at the origin in momemtum space, while the propagator is finite in the physical region. One way out of the impasse is to introduce a singularity in the nonperturbative coupling constant. Redefining the  potential in terms of this singular coupling  ($\sim 1/q^4$ ),  I will analyze in detail two cases for which I am able to reproduce both the propagator and the potential: i)  Dyson-Schwinger OGE  and ii)  mG + Dyson-Schwinger OGE. Finally, by softening the singularity I am able to fit a screened  interquark heavy meson potential \cite{Gonzalez:2011zc}.

The results of this investigation will be presented as follows. In section 2, I introduce the formalism and discuss the changes after incorporating the modified Gribov term in the propagator. In section 3, I study the corresponding potential.
In section 4, I introduce, as a consequence of mathematical reasonings, a singularity in the coupling constant and discuss its consequences. Section 5 is dedicated to the formulation of the screened potentials and I finish in section 6 by drawing some conclusions.

\begin{figure}[htb]
\begin{center}
\includegraphics[scale=1.0]{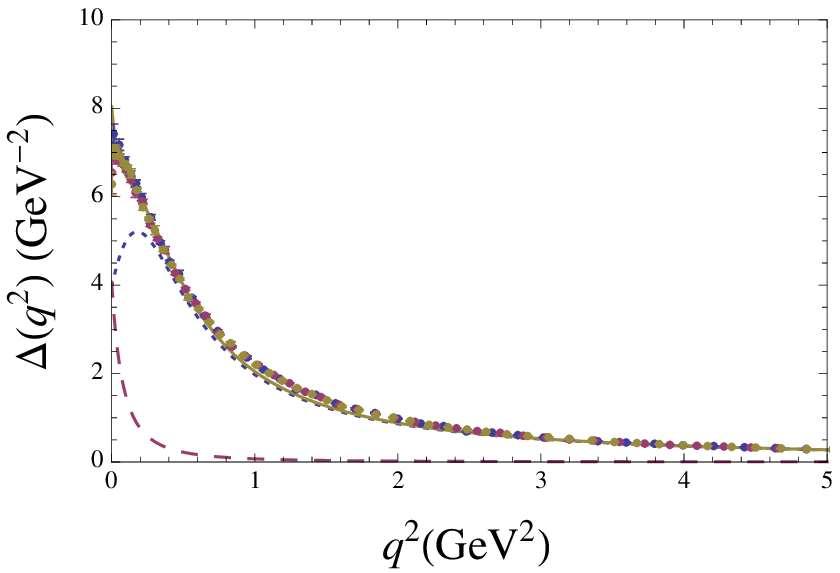}
 \vskip -5.6cm \hskip 4.4cm
\includegraphics[scale=0.45]{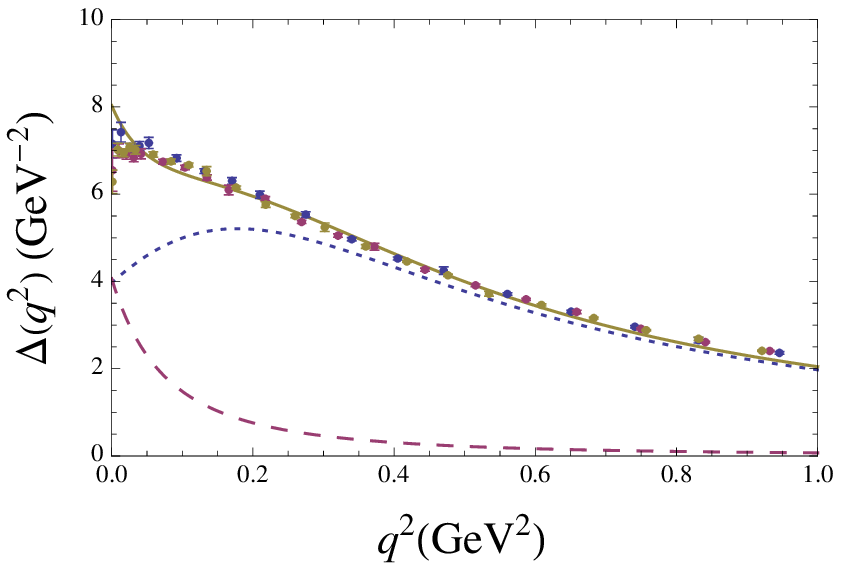}
\vskip 3cm
\caption{Fit to the lattice propagator data of ref. \cite{Bogolubsky:2007ud} using the logarithmic mass equation and the following set of parameters: a)  Dyson-Schwinger propagator (dashed)  $\Lambda = 0.300$ GeV, $\delta = 1/11$, $\rho =1.0$,  $m_0 = 0.50$ GeV, $\mu = 4.5 $ GeV,  $\rho_1= 1.0$, $c= 0.20$ ; b) Confining propagator (dotted)
$s= 1.044$ GeV$^2$, $m_c = 0.390$ GeV. The total propagator is represented by a continuous line. The inset reproduces the behavior close to the origin.}
\label{propagator1}
\end{center}
\end{figure}

\section{The Gluon Propagator}

Infrared finite solutions for the gluon propagator of quenched QCD are obtained from the
gauge-invariant non-linear Schwinger-Dyson equation formulated in the Landau gauge of
the background field method. These solutions may be fitted using a massive propagator.

 At the level of the Dyson-Schwinger equations (DSE)
the  generation of such a  mass is associated with 
the existence of 
infrared finite solutions for the gluon propagator, $\Delta (q^2)$,
i.e. solutions with  $\Delta^{-1}(0) > 0$.
Such solutions may  
be  fitted  by     ``massive''  euclidean propagator  of   the form 

\begin{equation}
 \Delta_{DS}^{-1}(q^2) =   m^2(q^2) + q^2 (1 + \Pi (q^2, \mu^2))
 \label{propagatorDS}
 \end{equation}
where $m^2(q^2)$ and $\Pi[q^2, \mu^2)$ depend non-trivially  on the momentum  transfer $q^2$.
One physically motivated possibility, which we shall use in here, is  the so called logarithmic mass running, which is defined by

\begin{equation}
m^2 (q^2)= m^2_0\left[\ln\left(\frac{q^2 + \rho m_0^2}{\Lambda^2}\right)
\bigg/\ln\left(\frac{\rho m_0^2}{\Lambda^2}\right)\right]^{-1 -\delta},
\label{rmass}
\end{equation}
where $m_0, \rho$ and  $\delta$ are parameters whose values are chosen to fit the lattice propagator and $\Lambda$ is the $QCD$ scale.

In order to fit the lattice data at a specific scale $\mu^ 2$ the following functional form  has been proposed \cite{Aguilar:2011yb},

\begin{equation}
\Delta_{DS}^{-1}(q^2)= m^2(q^2) + q^2\left[1+ c \ln\left(\frac{q^2 +\rho_1\,m^2(q^2)}{\mu^2}\right)\right].
\label{propagatorABP}
\end{equation}  

To this propagator I add an effective propagator of the form,

\begin{equation}
  s \Delta_{conf}^{-1}(q^2 )   =  (q^2 + m_c^ 2)^2, 
\label{mGribov}
\end{equation}
where $s$ is a dimensional constant and $m_c$ a mass determing the value of the full propagator at the origin. This behavior has been proposed  
to describe how confinement leads  to chiral symmetry breaking through a gap equation \cite{Cornwall:2010ap,Doff:2011sy}.
Thus our full propagator becomes
\begin{eqnarray}
\Delta^{\mu \nu} (q^2)  & = & \delta^{\mu \nu}  \Delta(q^2),  \nonumber\\
\Delta(q^2) &=& \Delta_{DS}(q^2) +  \Delta_{conf}(q^2 ).
\label{propagator}
\end{eqnarray}
 In Fig. \ref{propagator1}, I show how the lattice data can be fitted with this new propagator
and the functional forms for the DS gluon propagator presented  above \footnote{Note that in ref.  \cite{Aguilar:2011yb} other functional forms have been used.}.

\section{The Modified Dyson-Schwinger Potential}

Due to the presence of this dynamical gluon mass the strong effective
charge extracted from these solutions freezes at a finite value, giving rise to an infrared fixed
point for QCD. 
The  non-perturbative  generalization  of $\alpha(q^2)$
the  QCD  running  coupling, comes in the form
\begin{equation}
a(q^2) = \left[\beta_0 \ln \left(\frac{q^2 +\rho \, m^2(q^2)}{\Lambda^2}\right)\right]^{-1} ,
\label{alphalog}
\end{equation}
where  $a =\frac{\alpha}{4 \pi}$ and we take $\beta_0 = 11 - 2 n_f/3$ where $n_f$ is the number of flavors. 
The $m(q^2)$ in the argument of the logarithm 
tames  the   Landau pole, and $a(q^2)$ freezes 
at a  finite value in the IR, namely  
\mbox{$a^{-1}(0)= \beta_0 \ln (\rho \, m^2(0)/\Lambda^2)$} \cite{Cornwall:1982zr,Aguilar:2006gr}.

\begin{figure}[htb]
\begin{center}
\vskip 0.2cm
\includegraphics[scale=1.0]{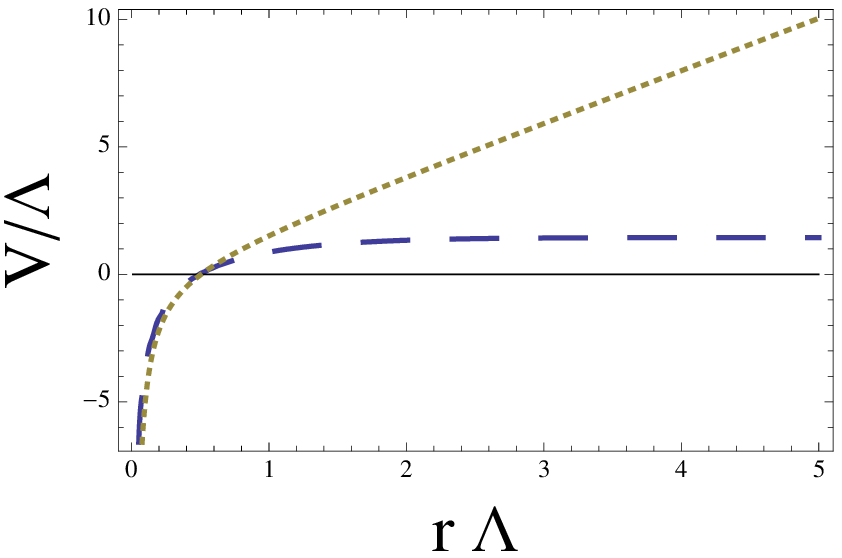}
\caption{The dashed curve shows the potential calculated using Eq.\ref{potential} with the parameters obtained by fitting   the lattice propagator  of ref. \cite{Bogolubsky:2007ud} using the full propagator Eqs. \ref{propagator} :  $\Lambda = 0.300$ GeV, $\delta = 1/11$, $\rho=1.0$,  $m_0 = 0.50$ GeV, $s= 1.044$ GeV$^2$, $m_c = 0.390$ GeV and particularizing for $n_f =4$, i.e. $\beta_0= 25/3$ in Eq.\ref{alphalog}. The dotted curve represents the Cornell potential as given in ref. \cite{Gonzalez:2011zc}.}
\end{center}
\label{potential0}
\end{figure}

The potential between the heavy quarks is obtained from the  One Gluon Exchange  potential  defined from the following propagator, \cite{Gonzalez:2011zc}

\begin{equation}
\frac{4 \pi C_F a(q^2)}{q^2 + m(q^2)},
\end{equation}
where $C_F$ is the Casimir eigenvalue of the fundamental representation of SU(3) [$C_F=4/3$]. To this term we add the modified Gribov propagator in Eq. \ref{mGribov}.

 The potential between static charges is related to the Fourier transform of the time-time component of the  full gluon propagator which after some trivial algebra becomes.

\begin{equation}
V({\mathbf r})=	- \frac{1}{|\mathbf r|}\int_{0}^{\infty}\!\!\! d {|\mathbf q}|\;{|\mathbf q}|\,\left(8 C_F\frac{a( \mathbf q^2)}{\mathbf q^2 + m(\mathbf q^2)} + \frac{s}{(\mathbf q^2 + m_c)^2}\right)
\sin({|{\mathbf q}||{\mathbf r}|}).
\label{potential}
\end{equation}
For constant mass parameter the mG propagator has an exact Fourier transform leading to

\begin{equation}
V_{conf} (\mathbf{r}) = -\; \sigma\; \frac{e^{-r \;m_c}}{ 2 m_c },
\label{confpotential}
\end{equation}
where $\sigma= s/4 \pi$. 

This potential for sufficiently small $m_c$ the potential behaves as linearly rising $V_{conf}(\mathbf{r) \sim }\sigma(  - \frac{1}{2 m_c} + \frac{r}{2} + \ldots)$. Do the data support a small enough mass parameter to define a (almost)-liner behavior for the values of $r$ required to fit the spectrum? In order to compare with the Cornell potential we have to implement the Sommer substraction as described in ref. \cite{Gonzalez:2011zc,Sommer:1993ce}.

In Fig. \ref{potential0}, I compare the potential obtained  from the fit to the lattice propagator of the previous section to the Cornell potential \cite{Eichten:1974af,Quigg:1979vr,Eichten:1979ms,Eichten:2007qx}. I note that the confining mass is  too large to produce a linear rise at the relevant $r$ values.
Therefore the mechanism described above does not  provide the required dynamics to describe quenched QCD confinement \cite{Greensite:2003xf}.

Let me now proceed in the opposite way. I fit the Cornell potential using Eq. \ref{potential}  and construct the corresponding propagator. It is clear from the fit that the Cornell potential requires $m_c$ very close to zero. What happens then to the propagator?

\begin{figure}[htb]
\vskip 0.2cm
\includegraphics[scale=0.82]{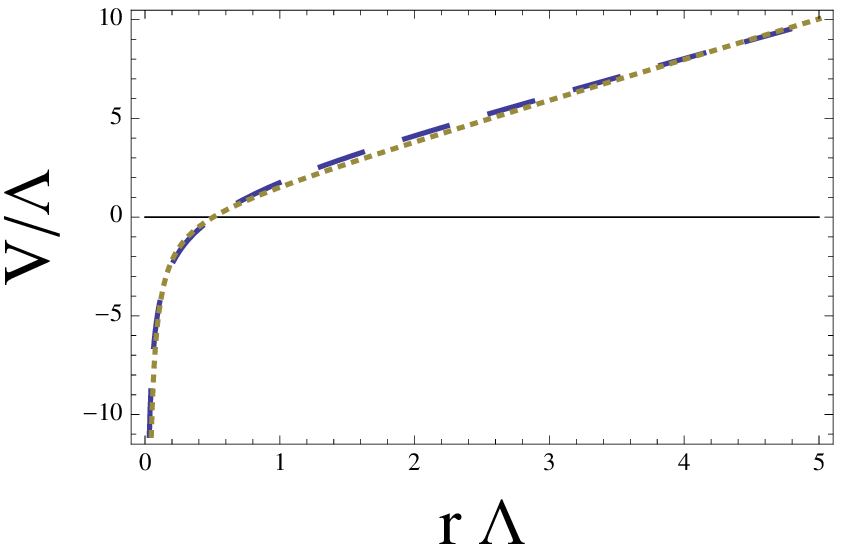}
\vskip -4.6cm \hskip 8.0cm
\includegraphics[scale=0.8]{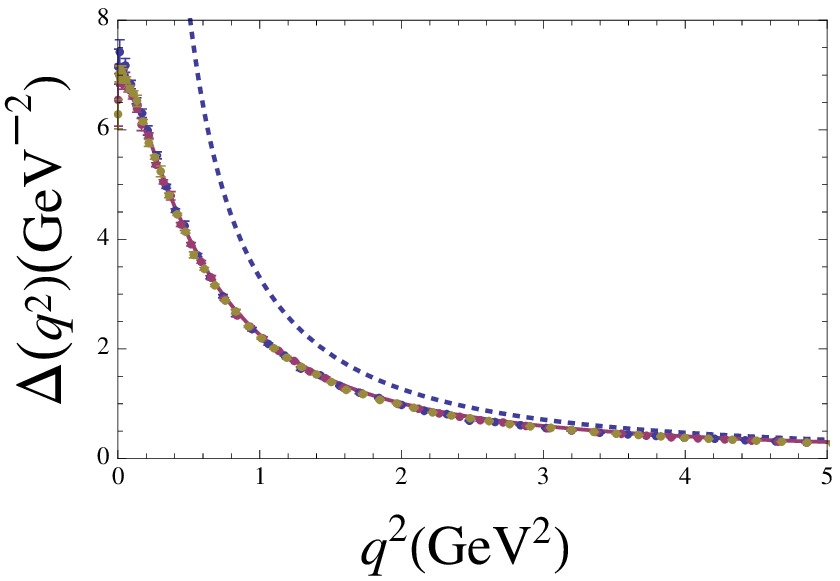}
\vskip 0.2cm
\caption{Left: Modified Dyson Schwinger potential (dashed) defined by the fllowing parameters: $\Lambda = 0.300$ GeV, $\delta = 1/11$, $\rho =2.0$,  $m_0 = 0.37$ GeV, $s= 4.316$ GeV$^2$, $m_c = 0.0$ GeV and $\beta_0= 25/3$. The dotted curve represents the Cornell potential as given in ref. \cite{Gonzalez:2011zc}. Right: The dotted curve represents the propagator obtained by using the parameters which fitted the Cornell potential. The continuous curve, which is almost indistinguishable from the data corresponds to the same fit without mG term, i.e. taking $s=0$. }
\label{potentialpropagador1}
\end{figure}

In Fig. \ref{potentialpropagador1},  I show the fit to the Cornell potential and  the corresponding propagator together with the lattice data. It is clear that the fitting of the Cornell potential requires a svery mall mass and a large coupling $s$, while the propagator requires a small coupling $s$ if we choose a small mass. I show in the figure also the extreme case with $s=0$, i.e.  only with OGE propagator, which turns out to give a  very precise fit to the propagator. Thus, one can fit the propagator with only the DS term  as was done in  ref. \cite{Aguilar:2011yb}. The corresponding parameters are those given in the figure caption. 

I am arriving to a impasse. The parameters which fit the propagator do not fit the potential and viceversa. The potential requires a confinement term \`a la Gribov while the propagator is finite at the origin.  I therefore conclude from the analysis that it is impossible to fit the propagator and the potential simultaneously, since the potential requires a Gribov singularity for quenched QCD.

\section{Confinement Coupling Constant}

How can we solve this puzzle? The way I forsee is dynamical. I need a singularity at the origin either  in the confining mG or in the DS propagators in order to reproduce the Gribov behavior. However,   the lattice data seem to imply that there is no singular behavior near the origin. One could still aim mathematically at a ``hidden" singularity  below the first data points. This solution does not make sense because a very narrow singularities in momentum space, becomes a fast flattening in coordinate space and therefore no linear rise will appear from such mechanism. My proposal here is  to change the coupling constant and incorporate there the singularity.

I have studied two mechanisms

\begin{itemize}  

\item[ i)] The Dyson-Schwinger scheme:  the coupling constant changes as,

\begin{eqnarray}
a_{total} (q^2) & = & a_{conf}(q^ 2) + a_{DS} (q^2), \nonumber \\
a_{conf} (q^2) & =& a_{conf}(0)\; \frac{\Lambda ^4}{q^4}, \nonumber \\
a_{DS} (q^2) & = & \left[\beta_0 \ln \left(\frac{q^2 +\rho \, m^2(q^2)}{\Lambda^2}\right)\right]^{-1}.\nonumber\\
\end{eqnarray}

This scheme has no mG term and the additional contribution could arise from a nonperturbative vertex correction.

\item[ ii)] The Gribov scheme: the  potential acquires an additional nonperturbative coupling  multiplying the modified Gribov propagator,

\begin{eqnarray}
\Delta_{conf} (q^2) &=& a_{conf} (q^2) \; \frac{ s }{(q^2 + m_c^2)^2}, \nonumber \\
 a_{conf} (q^2) &= & a_{conf}(0) \; \frac{	\Lambda ^4}{q^4},\nonumber\\
\label{fullpotentialEq}
\end{eqnarray}

\end{itemize}

\begin{figure}[htb]
\vskip 0.2cm
\includegraphics[scale=0.8]{propagador0S.eps}
\vskip -4.75cm \hskip 8.0cm
\includegraphics[scale=0.82]{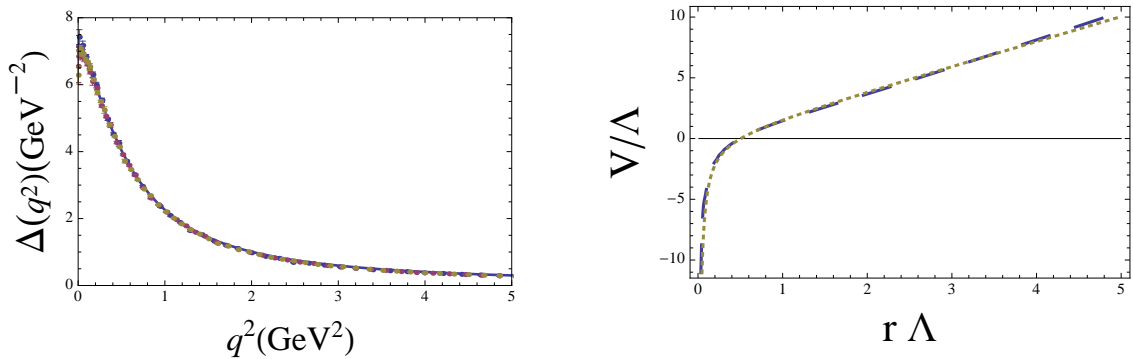}
\vskip 0.2cm
\caption{Left: Fit to the lattice propagator data of ref. \cite{Bogolubsky:2007ud} using the logarithmic mass equation and the following set of parameters:  Dyson-Schwinger propagator (dashed)  $\Lambda = 0.300$ GeV, $\delta = 1/11$, $\rho =2.0$,  $m_0 = 0.37$ GeV, $\mu = 4.5 $ GeV,  $\rho_1= 20.0$, $c= 0.27$.  Right: The dashed curve shows the potential calculated using Eq.\ref{potentialG} with the parameters obtained by fitting  the lattice propagator, particularizing for $n_f =4$, i.e. $\beta_0= 25/3$ in Eq.\ref{alphalog} and using $a_{conf(0)}  = 0.5$. The dotted curve represents the Cornell potential as given in ref. \cite{Gonzalez:2011zc}.}
\label{potentialpropagatorS1}
\end{figure}

\begin{figure}[htb]

\vskip 0.2cm
\includegraphics[scale=0.80]{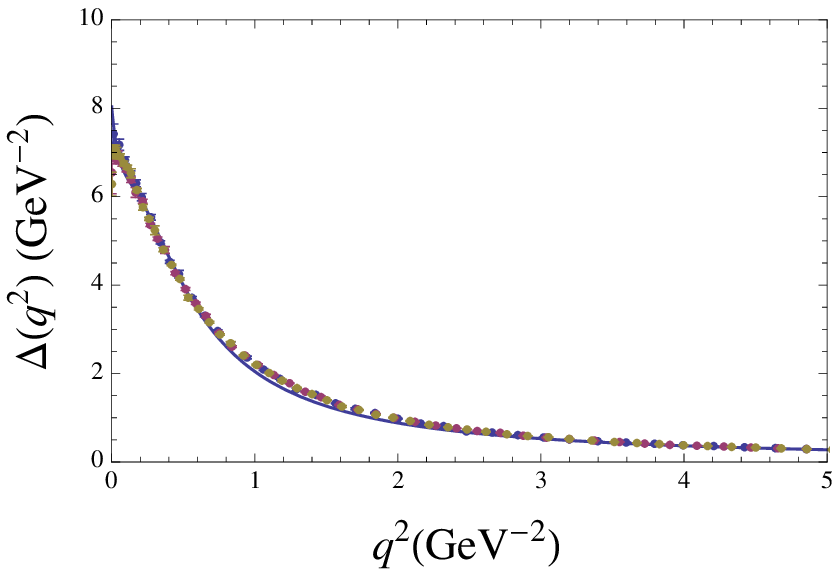}
\vskip -4.6cm \hskip 8cm
\includegraphics[scale=0.80]{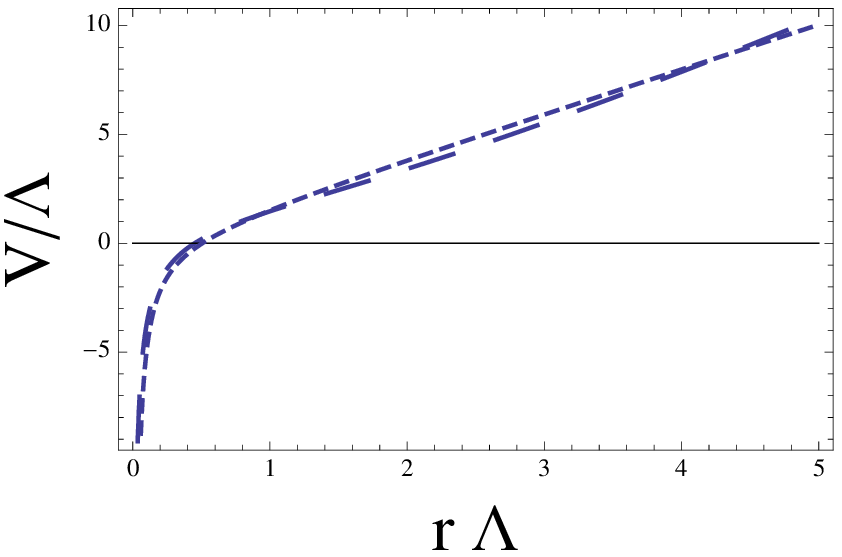}
\vskip 0.2cm
\caption{Left: Fit to the lattice propagator data of ref. \cite{Bogolubsky:2007ud} using the logarithmic mass equation and the following set of parameters: a)  Dyson-Schwinger propagator (dashed)  $\Lambda = 0.300$ GeV, $\delta = 1/11$, $\rho =2.0$,  $m_0 = 0.37$ GeV, $\mu = 4.5 $ GeV,  $\rho_1= 1.0$, $c= 0.2$; b) Confining propagator (dotted)
$s= 0.094$ GeV$^2$, $m_c = 0.390$ GeV.   Right: The dashed curve shows the potential calculated using Eq.\ref{potentialG} with the parameters obtained by fitting  the lattice propagator, particularizing for $n_f =4$, i.e. $\beta_0= 25/3$ in Eq.\ref{alphalog} and using $a_{conf(0)} s= 18.2$ GeV$^2$. The dotted curve represents the Cornell potential as given in ref. \cite{Gonzalez:2011zc}.}
\label{potentialpropagatorG2}
\end{figure}

The idea behind these Ans\"atze is that the strength of the $1/q^4$ singularity,
needed at the origin to achieve a linearly rising potential, cannot come from the propagator, which is finite, and therefore must come from the vertex.

The potentials become now 

\begin{itemize}

\item[ i)]
\begin{equation}
V({\mathbf r})=	- \frac{1}{|\mathbf r|}\int_{0}^{\infty}\!\!\! d {|\mathbf q}|\;{|\mathbf q}|\,\left(8 C_F\frac{a_{total}( \mathbf q^2)}{\mathbf q^2 + m(\mathbf q^ 2)}\right)
\sin({|{\mathbf q}||{\mathbf r}|}).
\label{potentialG1}
\end{equation}

\item[ ii)]

\begin{equation}
V({\mathbf r})=	- \frac{1}{|\mathbf r|}\int_{0}^{\infty}\!\!\! d {|\mathbf q}|\;{|\mathbf q}|\,\left(8 C_F\frac{a( \mathbf q^2)}{\mathbf q^2 + m( \mathbf q^ 2)} + \frac{a_{conf}(\mathbf q^2)\; \sigma}{(\mathbf q^2 + m_c)^2}\right)
\sin({|{\mathbf q}||{\mathbf r}|}).
\label{potentialG}
\end{equation}
\end{itemize}
In Figs. \ref{potentialpropagatorS1}  and \ref{potentialpropagatorG2}, I show the result of the above procedure for the two cases.

\begin{figure}[htb]
\vskip 0.2cm
\includegraphics[scale=0.8]{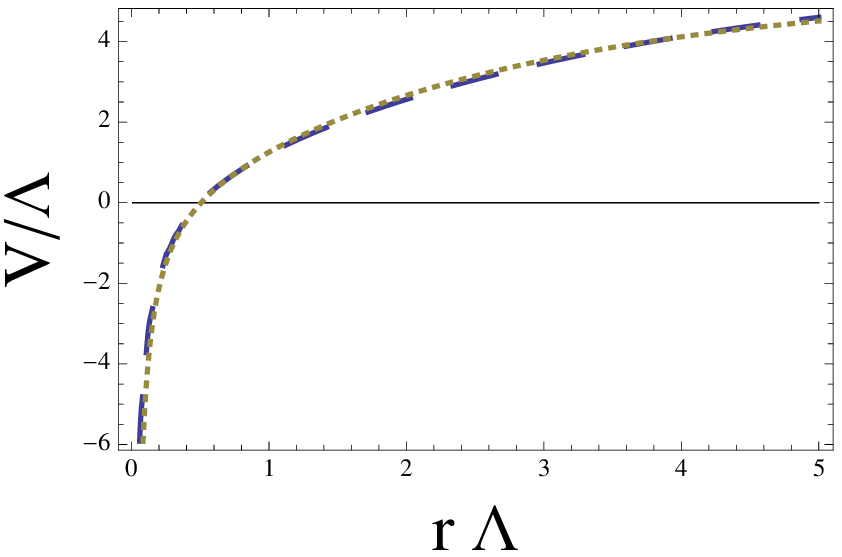}
\vskip -4.6cm \hskip 8cm
\includegraphics[scale=0.8]{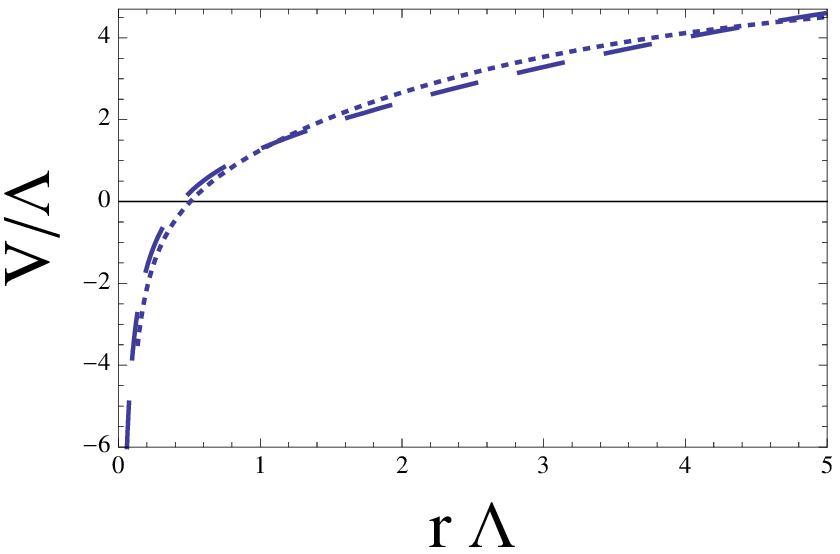}
\vskip 0.2cm
\caption{The dotted curve represents the Aachen potential as given in ref. \cite{Gonzalez:2011zc}. Left: Modified Dyson Schwinger with soft confinement potential (dashed) defined by the fllowing parameters: $\Lambda = 0.300$ GeV, $\delta = 1/11$, $\rho =2.0$,  $m_0 = 0.37$ GeV, $m_c = 0.390$ GeV, $a_{conf(0)}  = 0.5$,$m_a = 0.105$ GeV and $\beta_0= 25/3$. Right: Modified Gribov with soft confinement potential (dashed) defined by the fllowing parameters: $\Lambda = 0.300$ GeV, $\delta = 1/11$, $\rho =2.0$,  $m_0 = 0.378$ GeV, $a_{conf(0)} s= 18.2$ GeV$^2$, $m_c = 0.390$ GeV, $m_a = 0.090$ GeV and $\beta_0= 25/3$. }
\label{potential3S}
\end{figure}

The fits in the two cases are very good. In particular, for the DS case, the fit is almost perfect. I conclude that, as far as the propagator is finite, one way to achieve the dynamics of quenched QCD confinement in the scheme is by a singularities appearing in the vertex.

\section{Screened Potentials}

It is well known that in real QCD the coupling constant is finite \cite{Shirkov:1997wi}. In my scheme this can be  achieved avoiding the singularity by introducing a cutoff mass into the coupling constant, i.e. $q^2\rightarrow q^2 + m_a^2$.  From the point of view of interquark dynamics it is known that once the theory is unquenched, the linearly rising potential flattens at large $r$ leading to the so called Aachen potential \cite{Gonzalez:2011zc,Born:1989iv,Swanson:2005rc}. By using this cutoff mechanism I show the resulting potentials in Fig. \ref{potential3S}. Recall that the propagator fits are the same as before since the Gribov mechanism does not affect them. Again the fit is excellent in both cases, almost perfect in the DS case.

\section{Concluding Remarks}
The gluon propagator in the Landau gauge in quenched lattice QCD  is finite. This has been shown to be the case also for the resolution of truncated Dyson-Schwinger equations. This finiteness implies a limiting value for the corresponing interquark potential which does not correspond to the potential obtained from quenched QCD which is linearly rising. I simplest solution I forsee is to implemant a Gribov singularity in the nonperturbative coupling constant.  In my calculations all the parameters, except those related to the singularity, have been fixed to the lattice propagator and the correct potential arises  from adjusting the singularity. In this way I am able to reproduce the Cornell potential with great precision. The screened potential, corresponding to unquenched QCD, arises naturally by modifying the Gribov singularity with an additional mass parameter.  At this point we are not able to adscribe physical meaning to the parameters since they arise not from fundamental equations but from  parametrizations.  

The conclusion of this study is that one is able to reproduce the gluon propagator and the heavy interquark potential  if one is able to associate an interesting dynamical content to the nonperturbative coupling constant. A more fundamental study of this element from the point of view of nonperturbative QCD studies will shed some light in the confinement mechanism.

\section*{Acknowledgement}
I would like to thank illuminating discussions with Arlene Aguilar, Adriano Natale and Pedro Gonz\' alez.  This work has been partially funded by the Ministerio de Economía y Competitividad and EU FEDER under contract FPA2010-21750-C02-01, by Consolider Ingenio 2010
CPAN (CSD2007-00042), by Generalitat Valenciana: Prometeo/2009/129, by the
European Integrated Infrastructure Initiative HadronPhysics3 (Grant number
283286).


\begin{thebibliography}{60}

\bibitem{Cornwall:1982zr}
J.~M.~Cornwall,
Phys.\ Rev.\ D {\bf 26}, 1453 (1982). 

\bibitem{Aguilar:2006gr}
  A.~C.~Aguilar and J.~Papavassiliou,
  JHEP {\bf 0612}, 012 (2006)


\bibitem{Gonzalez:2011zc}
  P.~Gonzalez, V.~Mathieu and V.~Vento,
  Phys.\ Rev.\ D {\bf 84} (2011) 114008
  [arXiv:1108.2347 [hep-ph]].
  
\bibitem{Bogolubsky:2007ud}
 I.~L.~Bogolubsky, E.~M.~Ilgenfritz, M.~Muller-Preussker and A.~Sternbeck,
PoS {LATTICE}, 290 (2007).
  
\bibitem{Gribov:1999ui}
  V.~N.~Gribov,
  Eur.\ Phys.\ J.\ C {\bf 10} (1999) 91
  [hep-ph/9902279].


\bibitem{Cornwall:2010ap}
  J.~M.~Cornwall,
  Phys.\ Rev.\ D {\bf 83} (2011) 076001
  [arXiv:1011.3524 [hep-ph]].

\bibitem{Doff:2011sy}
  A.~Doff, F.~A.~Machado and A.~A.~Natale,
  Annals Phys.\  {\bf 327} (2012) 1030
  [arXiv:1106.2860 [hep-ph]].


\bibitem{Aguilar:2011yb}
  A.~C.~Aguilar, D.~Binosi and J.~Papavassiliou,
  JHEP {\bf 1201} (2012) 050
  [arXiv:1108.5989 [hep-ph]].



\bibitem{Sommer:1993ce}
  R.~Sommer,
  Nucl.\ Phys.\  B {\bf 411} (1994) 839
  [arXiv:hep-lat/9310022].

\bibitem{Eichten:1974af}
  E.~Eichten, K.~Gottfried, T.~Kinoshita, J.~B.~Kogut, K.~D.~Lane, T.~-M.~Yan,
  Phys.\ Rev.\ Lett.\  {\bf 34 } (1975)  369-372.
  
\bibitem{Quigg:1979vr}
  C.~Quigg, J.~L.~Rosner,
  Phys.\ Rept.\  {\bf 56}, 167-235 (1979).
  
\bibitem{Eichten:1979ms}
  E.~Eichten, K.~Gottfried, T.~Kinoshita, K.~D.~Lane, T.~-M.~Yan,
  Phys.\ Rev.\  {\bf D21 } (1980)  203.
  
 
\bibitem{Eichten:2007qx}
  E.~Eichten, S.~Godfrey, H.~Mahlke and J.~L.~Rosner,
  Rev.\ Mod.\ Phys.\  {\bf 80} (2008) 1161
  [arXiv:hep-ph/0701208].
 
\bibitem{Greensite:2003xf}
  J.~Greensite, S.~Olejnik,
  Phys.\ Rev.\  {\bf D67 } (2003)  094503.
  [hep-lat/0302018].
 
\bibitem{Shirkov:1997wi}
  D.~V.~Shirkov and I.~L.~Solovtsov,
  Phys.\ Rev.\ Lett.\  {\bf 79} (1997) 1209
  [hep-ph/9704333].

 
\bibitem{Born:1989iv}
  K.~D.~Born, E.~Laermann, N.~Pirch, T.~F.~Walsh, P.~M.~Zerwas,
  Phys.\ Rev.\  {\bf D40 } (1989)  1653-1663.

\bibitem{Swanson:2005rc}
  E.~S.~Swanson,
  J.\ Phys.\ G {\bf 31} (2005) 845
  [arXiv:hep-ph/0504097].
 
\end{thebibliography}
\end{document}